\documentclass[aps,prx,twocolumn,amsmath,amssymb,superscriptaddress,longbibliography]{revtex4-2}
\usepackage[breaklinks,colorlinks,bookmarks=false,citecolor=blue,linkcolor=red,urlcolor=blue]{hyperref}
\usepackage{blindtext,enumitem,graphicx,dcolumn,color,xcolor,amsmath,braket,bm,changes,chemformula,cleveref}
\usepackage{ulem}
\graphicspath{{figures/}}

\newcommand{\mr}{moir\'e~}

\newcommand{\wsetwo}{WSe$_2$~}

\crefname{equation}{Eq.}{Eqs.}
\Crefname{equation}{Eq.}{Eqs.}
\crefname{figure}{Fig.}{Figs.}
\Crefname{figure}{Fig.}{Figs.}
\crefname{section}{Sec.}{Secs.}
\Crefname{section}{Sec.}{Secs.}

\begin{document}
\title{Tunable Stripe Order and Weak Superconductivity in the Moir\'{e} Hubbard Model}

\author{Alexander Wietek}
\email{awietek@flatironinstitute.org}
\affiliation{Center for Computational Quantum Physics, Flatiron Institute, 162 5th Avenue, New York, New York 10010, USA}
\affiliation{Max Planck Institute for the Physics of Complex Systems, N\"othnitzer Strasse 38, Dresden 01187, Germany}

\author{Jie Wang}
\affiliation{Center for Computational Quantum Physics, Flatiron Institute, 162 5th Avenue, New York, New York 10010, USA}

\author{Jiawei Zang}
\affiliation{Department of Physics, Columbia University, 538 W 120th Street, New York, New York 10027, USA}

\author{Jennifer Cano}
\affiliation{Center for Computational Quantum Physics, Flatiron Institute, 162 5th Avenue, New York, New York 10010, USA}
\affiliation{Department of Physics and Astronomy, Stony Brook University, Stony Brook, New York 11794, USA}

\author{Antoine Georges}
\affiliation{Center for Computational Quantum Physics, Flatiron Institute, 162 5th Avenue, New York, New York 10010, USA}
\affiliation{Coll\`ege de France, 11 place Marcelin Berthelot, 75005 Paris, France}
\affiliation{CPHT, CNRS, \'Ecole Polytechnique, Institut Polytechnique de Paris, Route de Saclay, 91128 Palaiseau, France}
\affiliation{DQMP, Universit\'e de Gen\`eve, 24 Quai Ernest Ansermet, CH-1211 Gen\`eve, Switzerland}

\author{Andrew Millis}
\affiliation{Center for Computational Quantum Physics, Flatiron Institute, 162 5th Avenue, New York, New York 10010, USA}
\affiliation{Department of Physics, Columbia University, 538 W 120th Street, New York, New York 10027, USA}

\begin{abstract}

The \mr Hubbard model describes correlations in certain homobilayer twisted transition metal dichalcogenides. Using exact diagonalization and density matrix renormalization group methods, we find magnetic Mott insulating and metallic phases, which, upon doping exhibit intertwined charge and spin ordering and, in some regimes, pair binding of holes. The phases are highly tunable via an interlayer potential difference.  Remarkably, the hole binding energy is found to be highly tunable revealing an experimentally accessible regime where holes become attractive.  In this attractive regime, we study the superconducting correlation function and point out the possibility of weak superconductivity.

\end{abstract}

\maketitle

\section{Introduction}
\begin{figure}[t]
    \centering
    \includegraphics[width=\columnwidth]{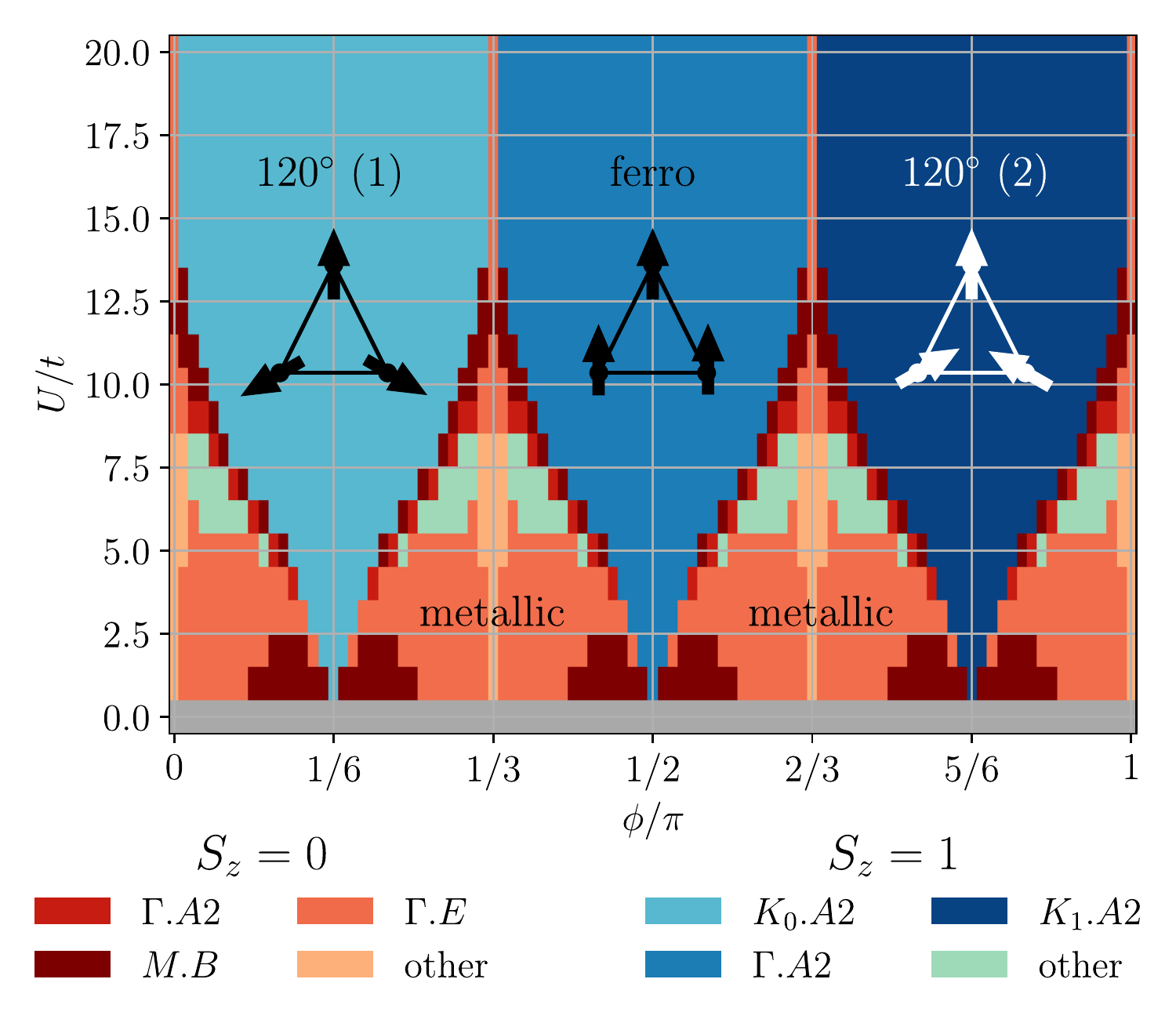}\\
    \includegraphics[width=0.35\columnwidth]{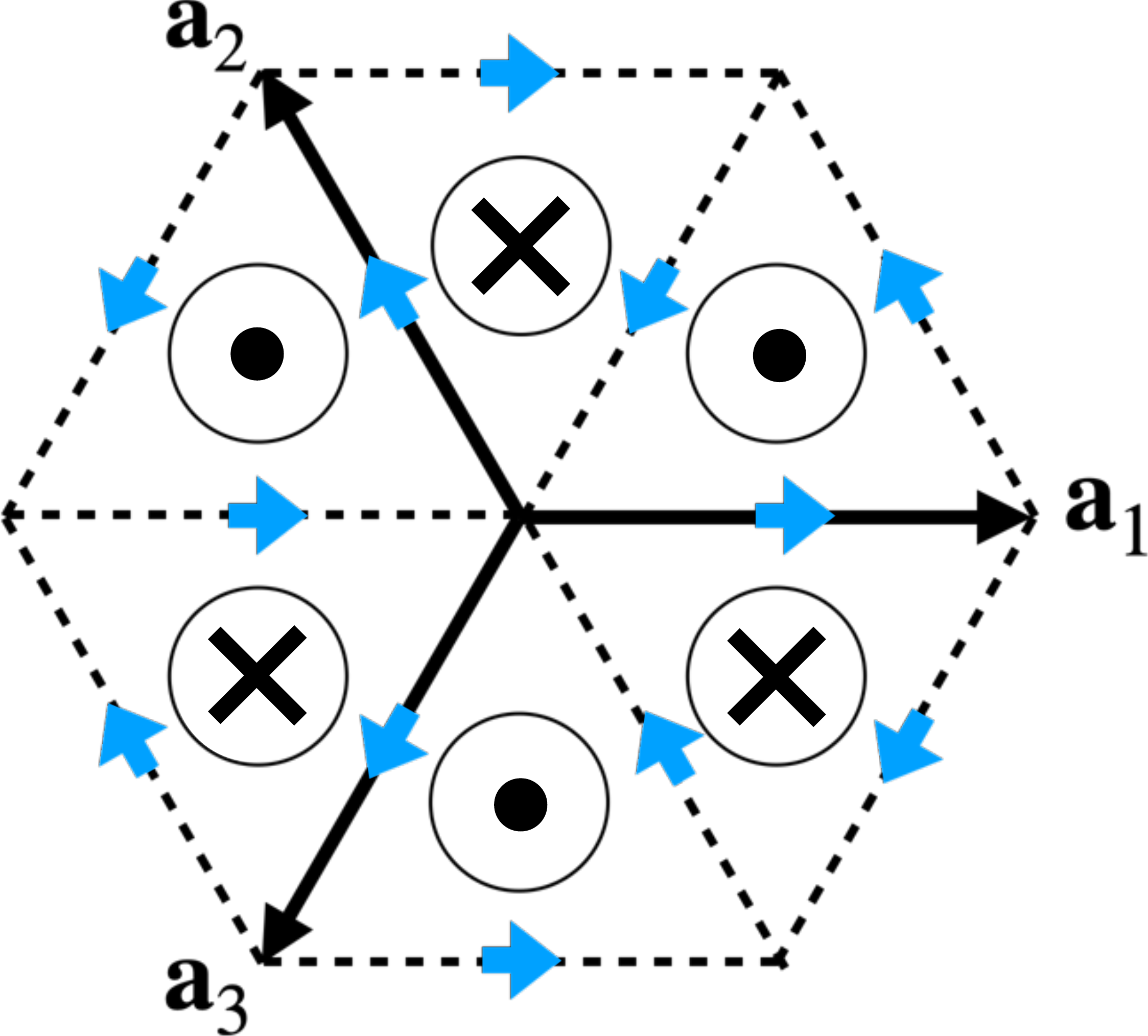} \hspace{1cm}
    \includegraphics[width=0.35\columnwidth]{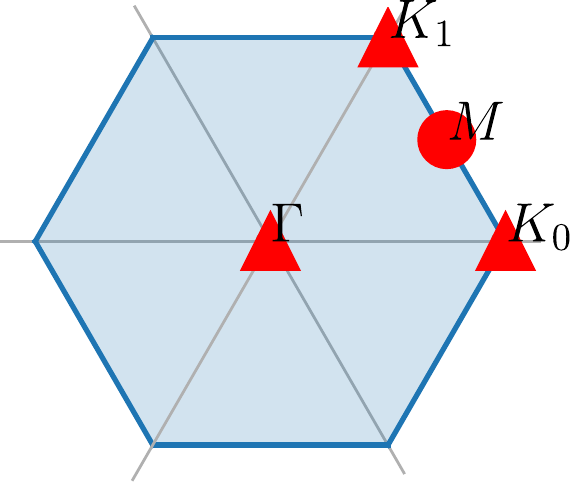}
    \caption{Approximate ground state phase diagram at half-filling on the $N_s=12$ cluster from Exact Diagonalization. The colors denote different quantum numbers of the first excited state. Red (Blue) colors correspond to $S_z=0$ ($S_z=1$) states at different momenta and point group symmetry representations, see \cref{sec:edgeometry}. The magnetic insulating regime at large-$U$ extends close to $U=0$ for $\phi=\pi/6, \pi/2, 5\pi/6$. Two $120^\circ$ N\'eel ordered states in the x-y plane with opposite chirality are realized as well as a x-y ferromagnet. The flux pattern of the \mr Hubbard model and the high symmetry momenta in the Brillouin zone are shown
    bottom left and right.}
    \label{fig:first_exc}
\end{figure}

Twisted \mr materials have attracted intense recent attention due to the range of correlated phenomena they exhibit and their versatile experimental tunability. Exotic emergent phenomena including correlated insulating states \cite{Tang:2020aa,Wang:2020aa}, quantum critically and tunable metal-insulator transitions (MIT) \cite{Li:2021aa,ghiotto2021quantum} have been recently realized in the twisted \wsetwo system, a typical class of transition metal dichalcogenides. The low energy physics of twisted \wsetwo is well captured by the so-called \mr Hubbard model, a variant of the standard triangular lattice Hubbard model in which the electron hopping amplitude acquires a spin-dependent phase~\cite{DasSarma_tTMD_PRR,Zang2021,Jie_StaggeredField,Kennes_tTMD_spinliquid,Zang_DMFT2021,Yao_RGflowSC}. The Hamiltonian $H$ of the \mr Hubbard model is the sum of  kinetic ($H_0$) and interaction ($H_I)$ terms with
\begin{equation}
H_0 = -t\sum_{\sigma=\uparrow,\downarrow}\sum_{\bm r,j=1,2,3}
 \left(e^{i\sigma\phi}c^\dagger_{\bm r+\bm a_j,\sigma} c_{\bm r,\sigma}  + \text{h.c.}
 \right),\label{eq:hamiltonian}
\end{equation}
and $H_I = U\sum_{\bm r}n_{\bm r\uparrow} n_{\bm r\downarrow}$ is the standard onsite repulsive Hubbard interaction. In Eq.~\ref{eq:hamiltonian}, $\bm a_{1,2}$ are primitive lattice vectors of the \mr triangular lattice with relative angle $2\pi/3$, $\bm a_3=-\bm a_1-\bm a_2$ shown in \cref{fig:first_exc}, and $\sigma=\uparrow,\downarrow$ represents the spin of the electron. 

The new feature of the model is the spin-dependent phase $\phi$ in hopping amplitude, $te^{i \sigma \phi}$. Initially obtained from DFT calculations~\cite{Wang:2020aa}, it captures the effect of the displacement field (gate voltage difference) between two layers. The displacement field  breaks the inversion symmetry between two layers. The introduced phase is the simplest way of breaking inversion symmetry, which breaks the $C_6$ rotation symmetry of the original model down to $C_3$,  and the fact that the phase is spin-dependent guarantees the time-reversal symmetry. 
$\phi$ produces a flux of $3\phi\sigma$ per triangular plaquette, alternating in sign between adjacent triangles and opposite for spin up and spin down.  Since a flux of $2\pi$ per triangular plaquette is equivalent to zero flux, the model is invariant under $\phi\rightarrow \phi+2\pi/3$. Further, a flux of $\pi$ $\equiv 3\pi$ per triangle corresponds to changing the sign of the hopping along each bond, equivalent to a particle-hole transformation. Thus the spectrum of the model at density $n$ and phase $\phi$ is the same as at density $2-n$ and flux $\phi \pm \pi/3$~\cite{DasSarma_tTMD_PRR,Zang2021,Jie_StaggeredField,Zang_DMFT2021}.

Importantly, both the carrier concentration and the phase are greatly tunable experimentally by the voltages associated with each layers. The sum of the two layer voltages determines the chemical potential while their difference, {\it i.e.} the displacement field, tunes the phase $\phi$~\cite{Wang:2020aa,DasSarma_tTMD_PRR,Zang2021,Jie_StaggeredField,Zang_DMFT2021}. In tWSe$_2$, physically achievable values of the displacement field correspond to changes of $\phi$ over the range of $-\frac{\pi}{3} \lesssim \phi \lesssim  \frac{\pi}{3}$. 

In this work, we explore the physics of the \mr Hubbard model along the aforementioned experimental tunable degrees of freedom: doping and displacement field. Through a combined exact diagonalization (ED) and density matrix renormalization group (DMRG) study, we obtain unambiguous numerical results that provide insight into the physics of the \mr Hubbard model. Key results include an approximate phase diagram at half-filling, carrier pairing, and indications of superconductivity away from half-filling at non-zero displacement fields.

\section{Phase diagram at half-filling}
To establish the ground state physics of the model \cref{eq:hamiltonian} at half-filling ($N=N_s$, where $N_s$ is the number of lattice sites) as a function of the interaction strength $U/t$ and the flux $\phi$, we employ ED on finite periodic lattices~\cite{Wietek2018}. We focus on the $N_s=12$ cluster, since this cluster is highly symmetric, can stabilize $120^\circ$ degree magnetic orders, and also features the $M$ point in the Brillouin zone, which we find to be important for the low-energy physics of the system. The next larger cluster featuring both the $K$ and $M$ point while having a $C_3$ symmetry would have $N_s=36$, which is not within reach of ED.

The upper panel of \cref{fig:first_exc} shows the phase diagram obtained from ED, along with color-coded indications of the momentum, point group representation, and the total magnetization $S_z$. In symmetry breaking phases, the quantum numbers of the low-lying ``tower-of-states'', which become the degenerate ground states in the thermodynamic limit, can be computed from group representation theory, Ref.~\cite{Wietek2016b}. 

\begin{figure*}
    \centering
    \includegraphics[width=\textwidth]{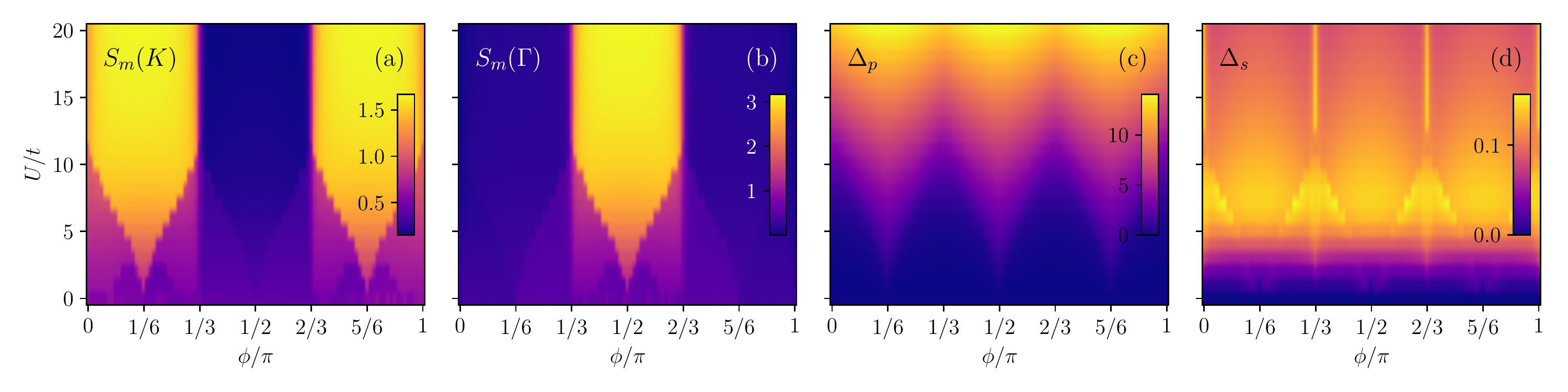}
    \caption{Magnetic order and gaps of the $N_s=12$ cluster from Exact
    Diagonalization. The magnetic structure factor $S_m(\mathbf{q})$ evaluated
    at points $\mathbf{q}=K$ (a) and Brillouin zone center $\mathbf{q}=\Gamma$ (b), indicating $120^\circ$ N\'eel order and ferromagnetic order in the $x$-$y$ plane, respectively. We observe a discontinuity as small values of $U/t$ around special values of $\phi=\pi/6, \pi/2, 5\pi/6$. (c) The single-particle gap $\Delta_p$ indicates the insulating and metallic regimes. (d) An enhanced spin gap $\Delta_s$ is observed close to the transition from the magnetic insulator to the metallic regime, possibly indicating non-magnetic insulating states.}
    \label{fig:edorders}
\end{figure*}

To complement this characterization of ground state phases we computed the ground state magnetic structure factor (in the $x$-$y$ plane),
\begin{equation}
    \label{eq:sofq}
    S_m(\mathbf{q}) = \frac{1}{N}\sum_{i,j=1}^N \text{e}^{i\bm{q} \cdot (\bm{r}_i - \bm{r}_j)} \braket{S^x_i S^x_j + S^y_i S^y_j},
\end{equation}
as well as the single-particle gap $\Delta_p$,
\begin{equation}
    \label{eq:deltap}
    \Delta_p = E_0(m+1,m) + E_0(m-1,m) - 2E_0(m,m),
\end{equation}
and the spin gap $\Delta_s$,
\begin{equation}
    \label{eq:deltas}
    \Delta_s = E_0(m+1,m-1) - E_0(m,m),
\end{equation}
where $E_0(m,n)$ denotes the ground state energy of the system with a number of $m$ electrons with spin up, and $n$ electrons with spin down. The results as a function of $U/t$ and $\phi$ are shown in \cref{fig:edorders}.

At large $U/t$, we find three different insulating regimes, with three different magnetic orders in the $x$-$y$ plane. For phase $\phi\in (\pi/3,2\pi/3)$ the spins orient ferromagnetically. This is evident in the large value of the magnetic structure factor at momentum $\Gamma\equiv(0,0)$ in
\cref{fig:edorders}(b). Consistent with this interpretation, the first excited state has discrete momentum, $\mathbf{k}=\Gamma=(0,0)$ and non-zero magnetization $S_z=1$, as expected for a translationally invariant state with ferromagnetic order. The insulating magnetic regimes at $\phi \in (0, \pi/3)$ and $\phi\in (2\pi/3, \pi)$  exhibit two peaks in the structure factor at $\mathbf{k}=K_0\equiv(4\pi/3,0)$ and $\mathbf{k}=K_1\equiv(2\pi/3, 2\pi / \sqrt{3})$. This is indicative of the planar $120^\circ$ N\'eel order, which can have two different chiralities. While the structure factor cannot distinguish between the two chiralities, the momentum of the first excitation indicates which chirality is realized, with the first excitation for $\phi \in (0, \pi/3)$ exhibiting excitation wave vector $\mathbf{k}=K_0$ while for $\phi\in (2\pi/3, \pi)$ yields momentum $\mathbf{k}=K_1$, which distinguishes the two chiralities.

In order to distinguish the metallic from the insulating regime, results for the single-particle gap $\Delta_p$ as defined in \cref{eq:deltap} are shown in \cref{fig:edorders}(c). We find that indeed the magnetically ordered regions coincide with regions of an enhanced single-particle gap. The magnetic insulating regimes extend close to $U/t=0$ for values of $\phi=\pi/6, \pi/2, 5\pi/6$.

These findings are consistent with previous results from a Hartree-Fock approximation~\cite{Zang2021}, where these three magnetic orders at large-$U$ as well as a metallic regime at small-$U$ were found. However, the MIT found here is shifted towards larger values of $U/t$ than the Hartree-Fock transition, consistent with recent dynamical mean-field results \cite{Zang_DMFT2021} and with more elaborate numerical results of the case  $\phi=0$, where the critical interaction strength for the MIT has been estimated to be $U_c/t\approx 8.7$~\cite{Wietek2021b}. 

Apart from the magnetic regimes \cref{fig:first_exc} shows several other regions with different quantum numbers of the first excitations, each of which could indicate a different phase being realized. At $\phi=0$, we find that in the intermediate coupling regime $9 \lesssim U/t \lesssim 11$ the lowest excited state has $S_z=0$, momentum $\mathbf{q}=\Gamma$ and, importantly, a point group representation E of the discrete dihedral point group $D_3$. This point group representation has been observed for chiral spin liquids~\cite{Wietek2017}. 
This observation would be consistent with recent proposals of a chiral spin liquid in the intermediate coupling regime at $\phi=0$~\cite{Szasz2020,Szasz2021,Chen2021}. This putative chiral spin liquid regime is suppressed by a finite displacement field, which is consistent with the expectation that breaking the SU(2) symmetry disfavors the formation of spin singlet and hence makes the spinon condensation harder to realize~\cite{Kennes_tTMD_spinliquid}.

At the particular values $\phi=\pi/6, \pi/2, 5\pi/6$ close to $U/t=0$ we observe several regimes where the first excitation has momentum $\mathbf{q} = M\equiv(\pi, \pi / \sqrt{3})$. The representation of the ground state is also different from the surrounding regime. Moreover, we observe a discontinuity in the structure factor in \cref{fig:edorders}(a,b). Therefore, this region could constitute a separate, likely metallic, phase, whose precise nature is yet to be determined. 

At the boundary of the magnetic insulators, we observe a state with momentum $\mathbf{q}=M$ and $S_z=0$, which could indicate a non-magnetic insulator. This would be also supported by the fact that the spin gap shown in \cref{fig:edorders}(d) is rather sizable in this region and exhibits a discontinuity. However, we cannot fully
exclude that this is an artifact of the finite cluster size, which could render this particular state to be the lowest excitation. Especially, we do not observe any particular feature in the magnetic structure factor in (a,b) in this regime.

\section{Physics at small hole doping}

\begin{figure*}
    \centering
    \includegraphics[width=\textwidth]{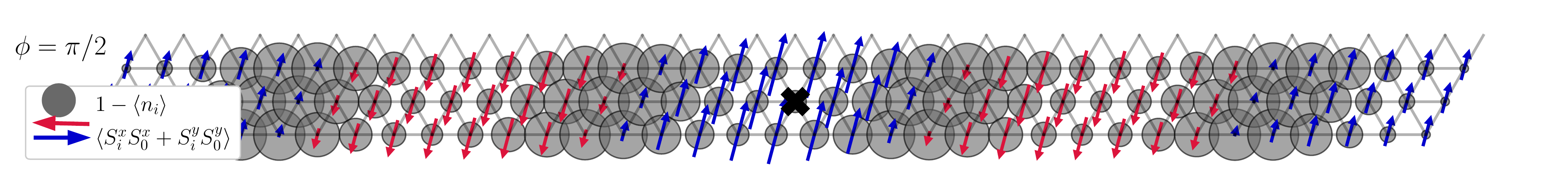}
    \caption{Hole-density $1-\braket{n_i}$ and spin correlation $\braket{S_i^xS_0^x + S_i^yS_0^y}$ of the ground state on the $36 \times 3$ YC3 cylinder for $U/t=8$ and small hole-doping $p\approx 0.074$, corresponding to $n_h=8$ holes at  $\phi=\pi/2$. The reference site of the spin correlations is marked with the black cross. We observe charge modulations with two holes per stripe. The spin correlation switches the sign at the maxima of the hole density, indicating intertwined spin and charge ordering.}
    \label{fig:dmrggroundstate}
\end{figure*}

\subsection{Intertwined spin and charge ordering}

Away from half-filling, charge and spin modulations of considerably larger wavelength than accessible by ED become important. We, therefore, apply the DMRG method to study the ground state properties of the system at small hole-doping on elongated cylindrical geometries. In this study, we focus on the YC3 geometry shown in~\cref{fig:dmrggroundstate}. The lattice is chosen to have periodic boundary conditions along the short cylinder length and open boundary conditions in the long direction, as conventionally chosen for use in matrix product state techniques.
The cylinder is well suited to study the ordered phases since it resolves the momenta $K_0$ and $K_1$. While YC4 and YC5 geometries of the triangular lattice Hubbard model at half-filling have been studied to unravel a chiral spin liquid~\cite{Szasz2020,Szasz2021}, these geometries do not resolve the $K$ points, and would hence introduce unphysical frustration. Similarly, twisted boundary conditions shift the resolved momenta. The YC6 cylinder would be a suitable candidate system. However, DMRG simulations at finite hole-doping in addition to the staggered magnetic field on the YC6 cylinder are currently beyond our reach.

We focus on the case $U/t=8$, which is believed to be relevant in twisted \ch{WSe2}~\cite{ghiotto2021quantum,Wang:2020aa}, where the system is metallic for $\phi=0$ while insulating for $\phi=\pi/6$ at half-filling. We pick particular values of $\phi=0,~\pi/6,~\pi/3$, and $\pi/2$. The case $\phi=0$ corresponds to the pure Hubbard model with nearest-neighbor hopping, where previous DMRG studies suggested the possibilities of pair-density wave~\cite{Peng2021} or chiral metal~\cite{Zhu2020} physics in this intermediate $U$ regime. At $\phi=\pi/6$ we are doping the system which according to \cref{fig:first_exc} exhibits $120^\circ$ N\'eel order in the $x$-$y$ plane at half-filling. For $\phi=\pi/2$ the system exhibits $x$-$y$ ferromagnetic order at half-filling. In contrast, at $\phi=\pi/3$ the system is metallic at half-filling.

We show ground state properties on the YC3 upon doping the ferromagnetic state at $\phi=\pi/2$ in \cref{fig:dmrggroundstate}. We observe a regular charge density modulation, where two holes form one stripe. The sign of the magnetic correlation switches on the maximum of the respective hole-density. Thus, the system exhibits typical intertwined spin and charge order. Originally, such orders were proposed by Hartree-Fock studies on the square lattice~\cite{Zaanen1989,Poilblanc1989,Machida1989,Kato1990} and have as of now been firmly established as the ground states in certain parameter regimes of the Hubbard model~\cite{LeBlanc2015,Zheng2017,Huang2017,Huang2018,Wietek2021}.


\begin{figure}
    \centering
    \includegraphics[width=\columnwidth]{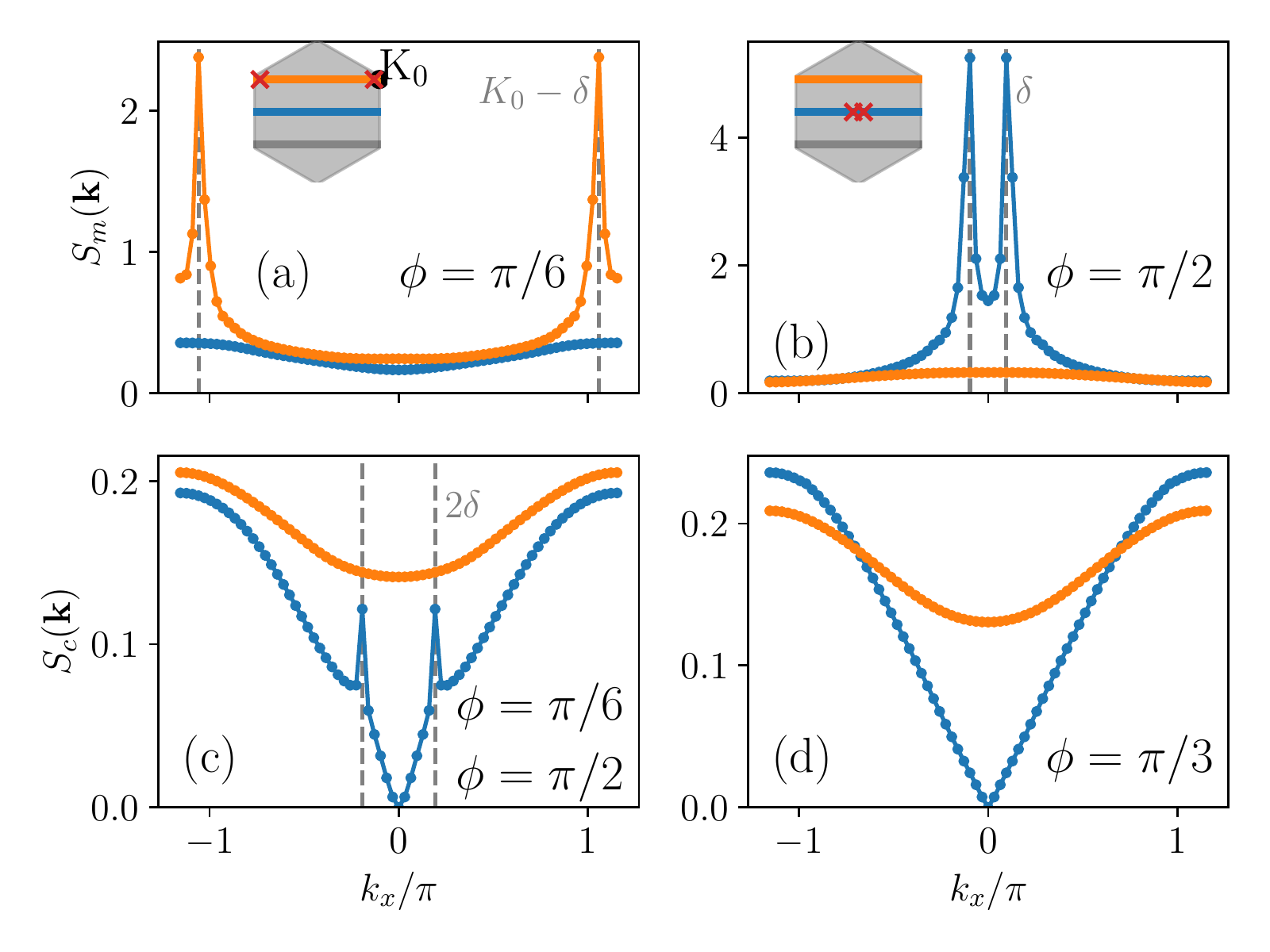}
    \caption{Intertwined charge and spin ordering on the $72\times 3$ YC3 cylinder at hole-doping $p=1/18$ and $U/t=8$. Blue and orange indicate the cut through the Brillouin zone shown in the insets. The peaks of spin structure factors in (a,b) are shifted by $\delta = \pi \sqrt{3} p$ from the ordering vectors $K_0$ (a) of the $120^\circ$ N\'eel order for $\phi=\pi/6$ and $\Gamma=(0,0)$ in (b) for $\phi=\pi/2$. The charge structure factor $S_c(\mathbf{k})$ in (c) is identical for $\phi=\pi/6$ and $\phi=\pi/2$ and is peaked at a wave vector $2\delta$, indicating stripe order. For $\phi=\pi/3$ in (d) no peak indicating stripe order is observed. The behavior $S_{\textrm{c}}(\bm{k}) \approx \alpha |k_x|$ at small $|k_x|$ indicates a metallic state.}
    \label{fig:charge_spin_sf}
\end{figure}

To quantify these observations, we computed the magnetic structure factor $S_m(\mathbf{k})$ and the charge structure factor,
\begin{equation}
  \label{eq:chargestructure}
  S_{\textrm{c}}(\bm{k}) = \frac{1}{N}\sum_{i,j=1}^N
  \text{e}^{i\bm{q}\cdot(\bm{r}_i - \bm{r}_j)}
  \braket{(n_i - \braket{n_i}) (n_j - \braket{n_j})},
\end{equation}
on the $72\times 3$ YC3 cylinder at hole-doping $p=1/18$ ($n_h=12$) for different values of $\phi$. \cref{fig:charge_spin_sf}(a) shows the magnetic structure factor $S_m(\mathbf{k})$ for $\phi=\pi/6$. We observed that its peak is shifted from the ordering vector $K_0$ of the $120^\circ$ N\'eel order by $\delta = \pi \sqrt{3} p = \pi \sqrt{3} (1 - n)$. Similarly, we observe a peak in the charge structure factor $S_c(\mathbf{k})$ at a wave vector of $2\delta$ in \cref{fig:charge_spin_sf}(c). Hence, stripe order where the wave length of the charge modulations is half the wave length of the spin modulation is also realized for $\phi = \pi/6$ and the spin modulation is a modulation of the $120^\circ$ order. To further verify the case of stripe ordering for $\phi = \pi/2$ shown in \cref{fig:dmrggroundstate}, $S_m(\mathbf{k})$ shown in \cref{fig:charge_spin_sf}(b) is similarly peaked at a small shifted wave vector $\delta$ instead of $\Gamma=(0,0)$, which would indicate uniform ferromagnetism. The charge structure factor \cref{fig:charge_spin_sf}(c) is identical for both $\phi=\pi/6$ and $\phi=\pi/2$, as expected from symmetry. Finally, at $\phi=\pi/3$ we do not observe any charge ordering, as can be seen from the structure factor $S_c(\mathbf{k})$ in \cref{fig:charge_spin_sf}(d). Instead, we clearly observe that 
\begin{equation}
  S_{\textrm{c}}(\bm{k}) \approx \alpha |k_x|,
\end{equation}
for small values of $|k_x|$. This is a key feature of a metallic state~\cite{Feynman1954,Capello2005,Tocchio2011,DeFranco2018}, as opposed to an insulating state which would be indicated by $S_{\textrm{c}}(\bm{k}) \approx \alpha k_x^2$. This agrees with the fact that the parent state at half-filling from ED in \cref{fig:first_exc} is already metallic.


\subsection{Effects of the flux $\phi$ on the hole-binding}

\begin{figure}
    \centering
    \includegraphics[width=\columnwidth]{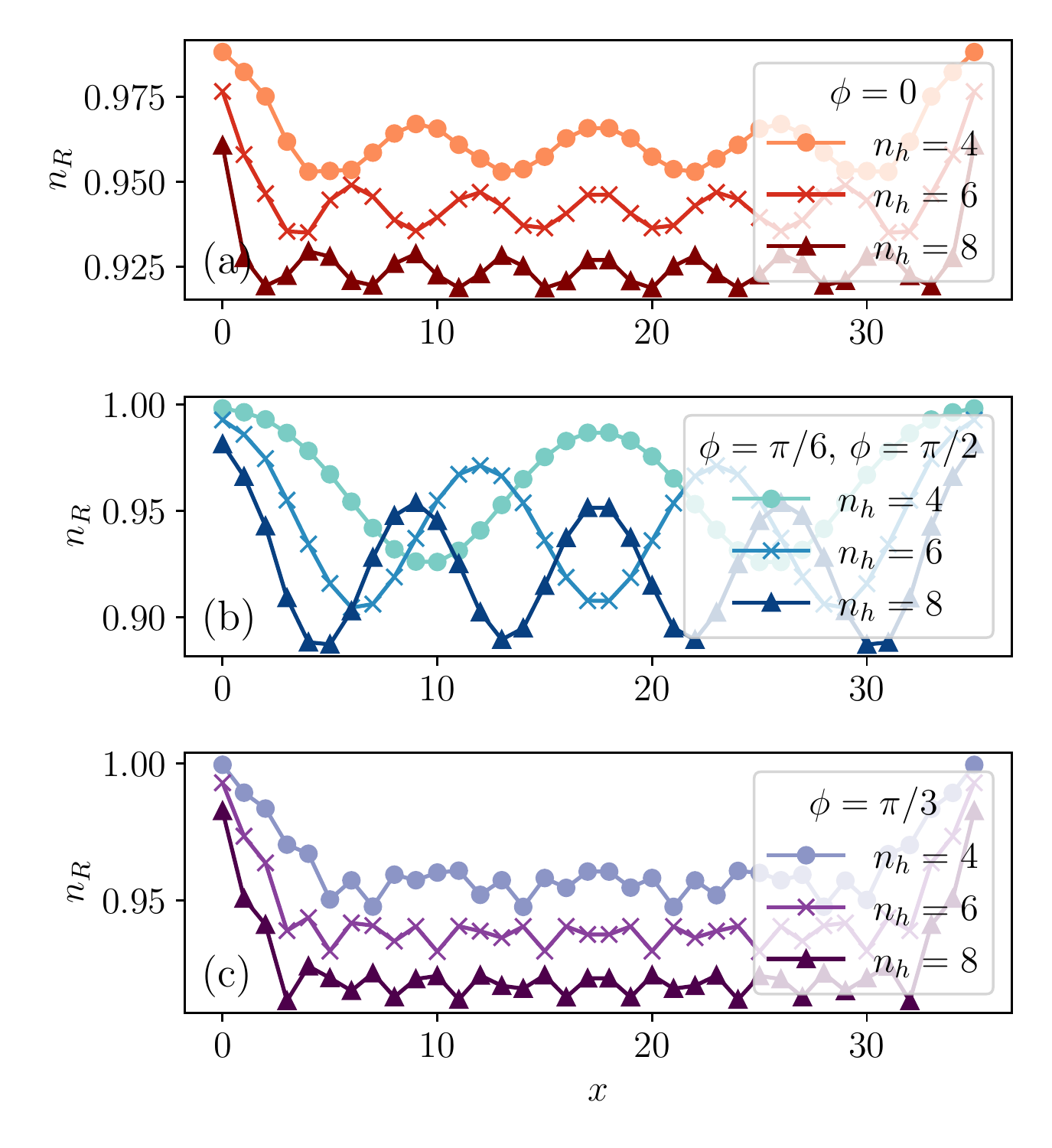}
    \caption{Rung-averaged density $n_R$ as a function of position $x$ at small
    hole-doping for $U/t=8.0$ in the case of (a) $\phi=0$, (b) $\phi=\pi/6, \pi/2$, and (c) $\phi = \pi/3$. The number of holes is denoted by $n_h$. The $\phi=\pi/6$ and $\phi=\pi/2$ yield identical density profiles as guaranteed by the symmetry discussed in the main text. This density profile indicate that single holes are separated at $\phi = 0$, while pairs of holes bind together to form a charge modulation at $\phi = \pi/6, \pi/2$. At $\phi=\pi/3$ no clear charge density wave patterns are formed.}
    \label{fig:density_profile}
\end{figure}

The charge structure factor $S_{\textrm{c}}(\bm{k})$ in \cref{fig:charge_spin_sf}(c,d) illustrates a non-trivial effect of the flux $\phi$ on the charge degrees of freedom upon hole-doping the parent states at half-filling. To further investigate the effect of $\phi$, we investigate the ground state electron density $n_R$ from DMRG for different values of $\phi$ in \cref{fig:density_profile} for $U/t=8$. Results are shown for hole-doping with $n_h=4, 6, 8$ holes and $\phi = 0, \pi/6, \pi/3$. The density profile of $\phi=\pi/6$ is identical to $\phi=\pi/2$ due to symmetry. For $\phi=0$ in \cref{fig:density_profile}(a) we find that the holes remain separated from one another. Hence, we observe one hole per stripe as previously reported in Ref.~\cite{Peng2021}. In contrast, at $\phi=\pi/6$ two holes bind together to form the charge modulation, so two holes per stripe are observed. Moreover, the hole-density at $\phi=\pi/3$ does not show regular charge density wave modulations, consistent with the absence of a peak in the $S_{\textrm{c}}(\bm{k})$ in  \cref{fig:charge_spin_sf}(d). Therefore, by changing the value of $\phi$ the system can be tuned from having repulsive, isolated holes to paired holes forming stripes to a more uniform charge density in a metallic state.

\begin{figure}[t]
    \centering
    \includegraphics[width=\columnwidth]{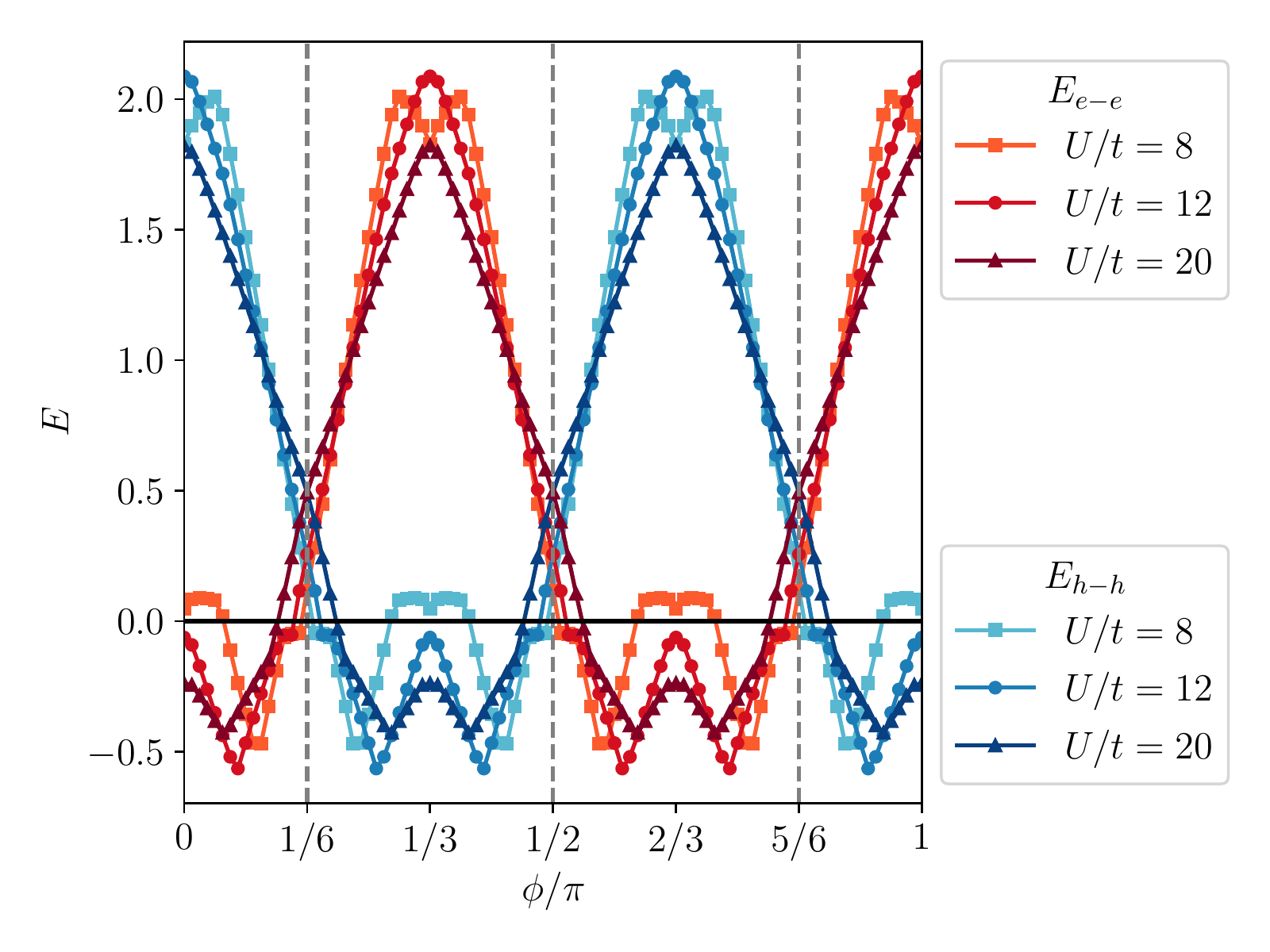}
    \caption{Hole-binding energy $E_{h-h}$ and electron binding energy $E_{e-e}$ as a function of $\phi$ evaluated on a $N_s=12$ for various values of $U/t$. Negative hole-binding energies favor the formation of bound hole pairs, consistent with the density profiles shown in \cref{fig:density_profile}. Electron- and hole-binding energies are related by symmetry.}
    \label{fig:binding_energies}
\end{figure}

To quantify this effect of the flux $\phi$ on the charge degrees of freedom, we investigate the hole-binding energy $E_{h-h}$ and the electron-binding energy $E_{e-e}$ defined by,
\begin{align}
  \begin{split}
  E_{h-h} \; = \; &E(m-1, m-1) +  E(m, m) \\
  &- E(m- 1, m) - E(m, m -1) 
\end{split}\\
  \begin{split}
  E_{e-e} \; = \;&E(m+1, m+1) +  E(m, m)\\
  &- E(m + 1, m) -  E(m, m+1),
  \end{split}
\end{align}
where $E(m, n)$ denotes the ground state energy in the sector with $m$ up- and $n$ down-electrons. The dependence of these energies on the $N_s=12$ cluster from ED for various values of $U/t$ is shown in \cref{fig:binding_energies}. The hole-binding energy $E_{h-h}$ is strongly positive at $\phi=0$ which implies that it is energetically more favorable to introduce two separate holes than a pair of holes, consistent with the isolated holes shown in \cref{fig:density_profile}(a). The value of $E_{h-h}$ decreases as a function of $\phi$, eventually becoming negative and attaining a minimum between $\phi=\pi/6$ and $\phi=\pi/3$. Negative hole-binding energies indicate that binding of two holes is energetically prefereable to having two isolated holes. The small hole-binding energy at $\pi/6$ is, thus, consistent with having bound hole pairs as shown in \cref{fig:density_profile}(b). Due to symmetry, the electron-binding energies $E_{e-e}$ are identical to the hole binding energies $E_{h-h}$ up to a shift of $\phi=\pi/3$.

\subsection{Weak superconductivity}

\begin{figure}[t]
  \centering
  \includegraphics[width=\columnwidth]{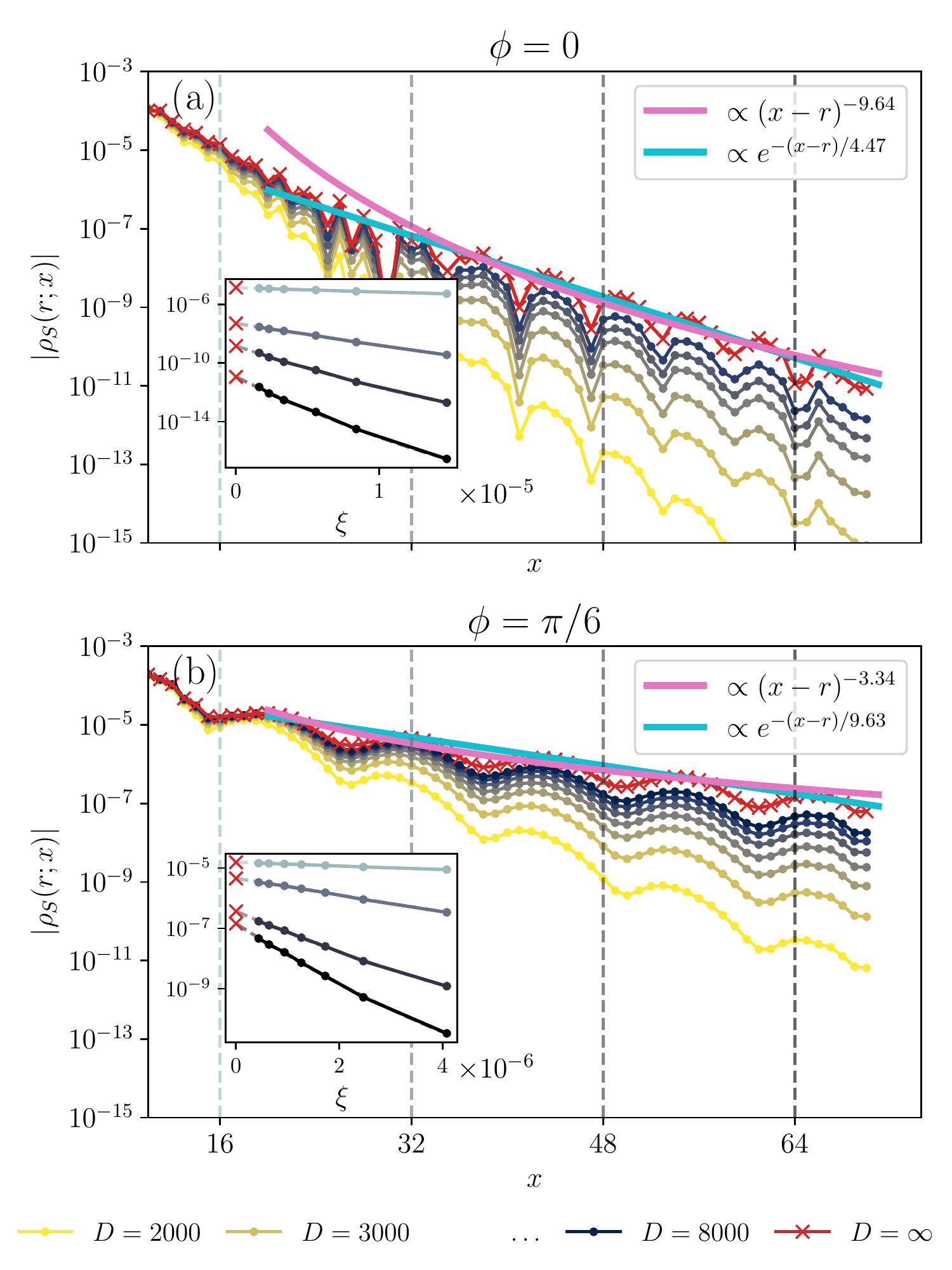}
  \caption{Singlet-pairing correlations
    $\rho_S(r,x) = \rho_S(\bm{r}, \alpha | \bm{x}, \beta)$ for $U/t=8$
    and $p=1/18$ on the $72\times 3$ cylinder as a function of
    $x=|\bm{r}_i - \bm{r}_j|$ for (a) $\phi = 0$ and (b)
    $\phi = \pi/6$. The reference point is chosen as $\bm{r}=(5,0)$
    and we choose $\alpha=\beta=(1,0)$. We extrapolate data from
    finite bond dimension $D$ to infinite bond dimension by fitting a
    second order polynomial to the truncated weight $\xi$ in DMRG, as
    shown in the insets. We fit both algebraic as well as exponential
    decay to the extrapolated correlations. The long-distance behavior
    of $\phi = 0$ is well-described by an exponential decay with
    correlation length $\approx 4.47$ while for $\phi = \pi/6$ both an
    algebraic decay with exponent $\approx 3.34$ and an exponential
    decay with correlation length $\approx 9.63$ can be fitted.}
  \label{fig:odlro}
\end{figure}

To determine whether the stripe state is accompanied by
superconductivity, we investigate the pairing properties of the
system. We focus on a particular set of parameters upon doping the
$120^\circ$ magnetically ordered phase, $U/t=8$, $p=1/18$ for both
$\phi=0$ and $\phi=\pi/6$. Superconductivity is diagnosed by two
means. First, we demonstrate off-diagonal (quasi) long range order in
the pairing correlations. We consider the singlet pairing matrix
$\rho_S(\bm{r}_i \alpha | \bm{r}_j \beta)$,
\begin{equation}
  \label{eq:singletpairingmatrix}
  \rho_S(\bm{r}_i,  \alpha| \bm{r}_k, \beta) = 
  \langle \Delta_{\bm{r}_i(\bm{r}_i + \alpha)}^\dagger \Delta_{\bm{r}_j(\bm{r}_j + \beta)}\rangle,
\end{equation}
where $\alpha,\beta$ denote the direction of the nearest-neighbor
lattice site on the triangular lattice and
\begin{equation}
  \label{eq:singletpairingop}
  \Delta_{\bm{r}_i\bm{r}_j}^\dagger= \frac{1}{\sqrt{2}}
  \left( 
    c^\dagger_{\bm{r}_i\uparrow} c^\dagger_{\bm{r}_j\downarrow} - 
    c^\dagger_{\bm{r}_j\uparrow} c^\dagger_{\bm{r}_i\downarrow}
  \right),
\end{equation}
denotes the singlet-pairing operator. On elongated quasi
one-dimensional geometries, long-range order even at $T=0$ is excluded
by the Mermin-Wagner theorem. Quasi long-range order, i.e. algebraic
decaying correlation functions, is interpreted as an indication of
true long-range order in the fully two dimensional system. We
performed ground state simulations for bond dimensions
$D=2000,\ldots,8000$ and extrapolated towards infinite bond
dimension. Results on the $72\times 3$ cylinder are shown in
\cref{fig:odlro}. The extrapolated correlation funciton for $\phi=0$
is well-described by an exponential decay with a correlation length of
$\approx 4.47$, shown in \cref{fig:odlro}(a). For $\phi=\pi/6$ the
correlation function is well described by either algebraic decay with
an exponent $\approx 3.34$ or an exponential decay with correlation
length $\approx 9.63$. It is, thus, difficult to discern whether
algebraic or exponential decay is realized based on the present data.

As compared to similar studies on different superconducting phases,
this exponent is rather large (which means the superconductivity
correlation is weak). Ref.~\cite{Gong2021}, for example, reported an
exponent $\eta \approx -0.96$ for the superconducting state realized
in the $t$-$J$ model on square lattice on a width $W=4$
cylinder. Hence, we interpret an exponent of $\eta=-3.34$ as a sign of
weak superconductivity.

\begin{figure}[t]
  \centering
  \includegraphics[width=\columnwidth]{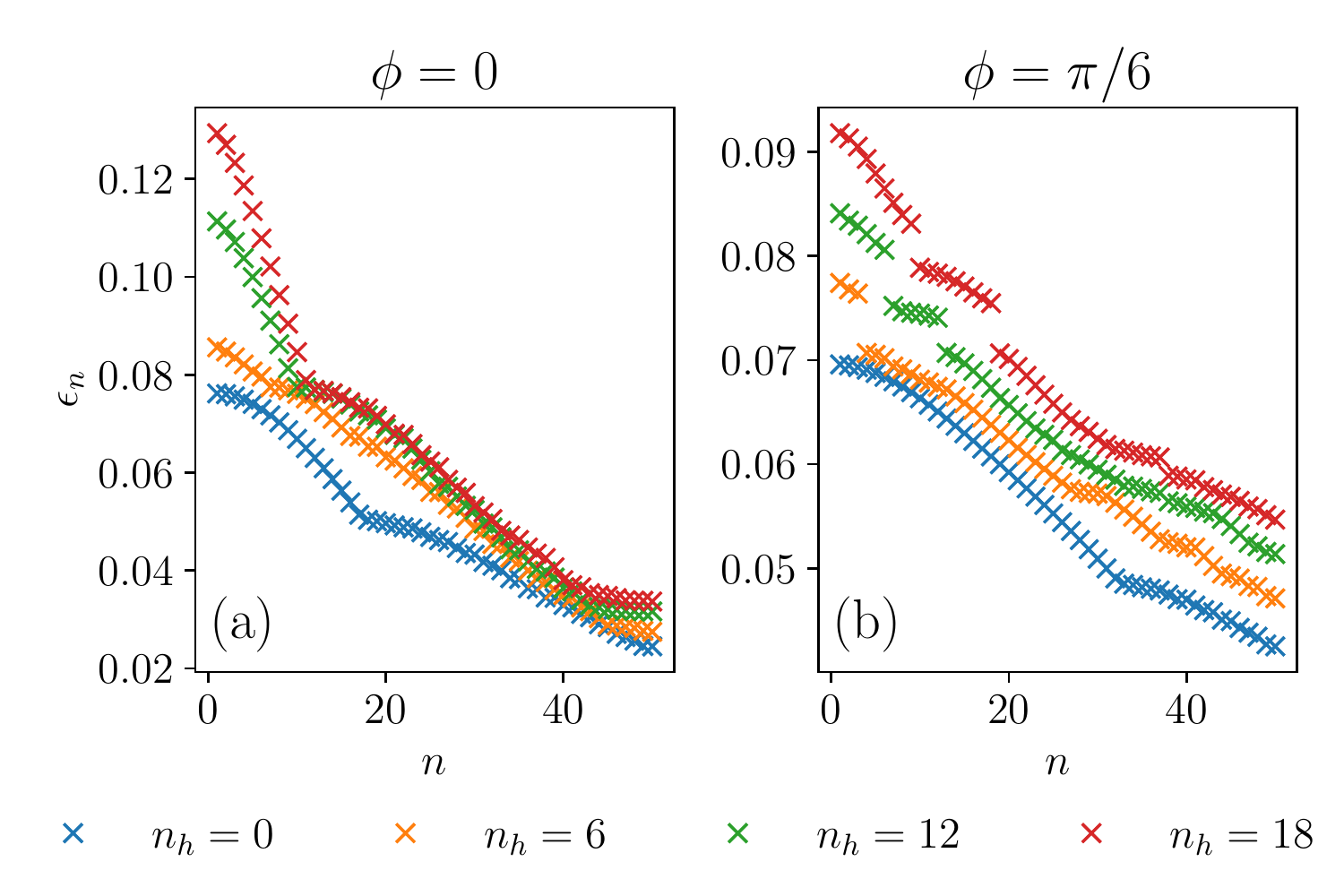}
  \caption{Spectrum of the singlet-pairing two-body density matrix as
    defined in \cref{eq:nonlocalmatrix} at $U/t=8$ on the
    $72 \times 3$ cylinder for (a) $\phi=0$ and (b) $\phi=\pi/6$. We
    compare different hole-dopings with $n_h=0,6,12,18$ and show the
    largest 50 eigenvalues only. For finite doping at $\phi=\pi/6$ in
    (b) a small gap to the continuum of residual eigenvalues is
    observed upon doping the system. The number of dominant
    eigenvalues equals the number of stripes, indicating a
    fragmentation by the stripes of the superconducting
    condensate. The absence of a large gap indicates only the
    possibility of a weak superconductivity. For $\phi=0$ in (a), we
    do not observe a gap in the eigenvalues and can, therefore,
    exclude superconducting order.}
  \label{fig:tbeigs}
\end{figure}

As a second means of diagnosing superconductivity, we investigate the
spectrum of the non-local singlet density matrix,
\begin{align}
  \label{eq:nonlocalmatrix}
  \begin{split}
    \hat{\rho}_S&(\bm{r}_i \alpha | \bm{r}_j \beta) \\
    &=\begin{cases} \rho_S(\bm{r}_i \alpha | \bm{r}_j \beta) &\mbox{if
      }
      \{ \bm{r}_i, \bm{r}_i + \alpha \} \cap \{ \bm{r}_j, \bm{r}_j + \beta \} = \emptyset \\
      0 &\mbox{else.}
    \end{cases}
  \end{split}
\end{align}

Upon Cooper pair condensation, we expect one or more eigenvalues to
become dominant over the residual continuum of eigenvalues~\cite{Wietek2022}. The spectra of
$\hat{\rho}_S(\bm{r}_i \alpha | \bm{r}_j \beta)$ for $U/t=8$, $\phi=\pi/6$, and $n_h=0,6,12,18$ on the $72 \times 3$ cylinder are shown in \cref{fig:tbeigs}. For $\phi=0$ in \cref{fig:tbeigs}(a) we
observe simply continuum of eigenvalues and no dominant eigenvalues,
indicating absence of superconductivity. Similarly, for $\phi=\pi/3$ at
half-filling $n_h=0$ shown in \cref{fig:tbeigs}(b), we observe a
continuum of eigenvalues. However, for $n_h=6$ we observe a small gap
of three dominant eigenvalues to the continuum of residual
eigenvalues.  Similarly, we observe six dominant eigenvalues for
$n_h=12$ and nine for $n_h=18$. Hence, the number of dominant
eigenvalues equals the number of stripes in the system. While this
phenomenon is analogous to the superconducting state in the
two-dimensional $t$-$J$ model on the square lattice, the gaps are
smaller in magnitude. Whereas, for the robust superconductivity in the
$t$-$J$ model on the square lattice gaps of the order of
$\Delta \approx 0.1$ have been observed~\cite{Wietek2022}, here we
report a gap of order $\Delta \approx 0.01$.  We again interpret this
small gap as a sign of weak superconductivity. However, the spectrum
in the hole-doped cases $n_h=6,12,18$ is clearly gapped in
\cref{fig:tbeigs}, as compared to the half-filled case or $\phi=0$
without superconductivity, which indicates a condensate of Cooper
pairs forming on the stripes of the system.

\section{Conclusion and Discussion}
The \mr Hubbard model is an effective description of the low-energy physics of twisted \ch{WSe2}, which features a spin-dependent staggered flux through the plaquettes of a triangular lattice. This model exhibits a rich phenomenology whose central aspects we have now established by our combined ED and DMRG study. We have established an approximate phase diagram at half-filling, where at larger values of $U/t$ two magnetic regimes with $120^\circ$ N\'eel order and one regime with xy-ferromagnetic order is realized. In comparison to the previous Hartree-Fock study~\cite{Zang2021}, the critical values of $U/t$ of the metal-insulator transition in shifted to larger values, in agreement with more elaborate studies on the pure triangular lattice Heisenberg model at $\phi=0$~\cite{Szasz2020,Wietek2021b}. At the particular values $\phi=\pi/6, \pi/2, 5\pi/6$ the magnetic insulating phases extend up to $U/t=0$ (within the numerical precision), which is also apparent from the single-particle gap in \cref{fig:edorders}(c). Since at $\phi=0$ at intermediate values of $8\lesssim U/t \lesssim 11$ the model has been shown to feature a non-magnetic insulating (possibly spin liquid) state, a natural question is where else to expect putative spin liquid regimes. The spin gap in \cref{fig:edorders}(d) is pronounced close to the metal-insulator transitions, which could be a first indication of a non-magnetic insulating state. However, further studies will be required to establish the phase diagram close to the metal-insulator transition at nonzero values of $\phi$.

Doping the parent magnetic insulating states leads to the formation of intertwined spin and charge ordering such that the wavelength of the charge modulations is half the wavelength of the spin modulation at $\phi=\pi/6$ and $\phi=\pi/2$. Interestingly, a nonzero flux $\phi$ leads to the formation of hole-pairs which we relate to a strong dependence of the hole-binding energies on $\phi$. While for $\phi=0$ individual holes are strongly repulsive, at intermediate and large values of $U/t$ the hole-binding energy is found to be negative, leading to an attractive force between the holes. Analogously, we find that a finite value of $\phi$ can enhance the superconducting pair correlation and lead to a gap in the eigenvalues of the two-body density matrix. However,  the pairing correlations at $U/t=8$ and $\phi=\pi/6$ are still weak and can be fitted by an algebraic decay with exponent $\approx 3.34$ or an exponential decay with correlation length $\approx 9.63$ in units of the lattice spacing, in either case too rapidly decaying to be consistent with a physical superconducting phase. Similarly, the gap in the spectrum of the two-body density matrix remains small. It will be interesting to determine how the superconductivity can be further enhanced, for example by further frustrating the magnetic order or tuning the value of the flux $\phi$.

Pair-density wave (PDW) orders have been previously discussed in particular parameter regimes of \cref{eq:hamiltonian}. For $\phi=0$ in the case of the pure triangular lattice Hubbard model we observe an alternating sign of the pairing correlation $\rho_S(r,x)$ in \cref{fig:odlro}(a), similar as has been observed in Ref.~\cite{Peng2021}. In our case, however, the absolute values $|\rho_S(r,x)|$ decay exponentially fast with a rather short correlation length. From this we cannot conclude that a PDW is realized at $U/t =8$ and $p=1/18$. At non-zero $\phi$, PDW order has also been suggested for small values of $U/t$ in DMRG~\cite{Venderley2019} and renormalization group~\cite{Wu2022} studies. For the  parameters studied in this manuscript, $U/t=8$ and $\phi = \pi/6, \pi/2$ we observe uniform pairing correlation inconsistent with a PDW state.

We note that, perhaps consistently with our finding of only weak superconductivity, no superconductivity has yet been observed in these materials, although superconductivity has been observed in the closely related twisted bilayer graphene materials \cite{Cao2017}. This is in interesting counterpoint to the high -$T_c$ cuprate materials, a square lattice material family in which robust superconductivity is observed. For the cuprates the accepted theoretical model is the square lattice Hubbard model. Whether the ground state in certain parameter regimes is superconducting is subject of ongoing research~\cite{Zheng2017} where an absence of superconductivity in the unfrustrated case has been noted~\cite{Qin2020}. However, in the closely related square-lattice $t$-$J$ model robust superconductivity was recently established~\cite{Jiang2020,Gong2021,HCJiang2021,Wietek2022}.

More generally, the great tunability of the \mr materials, in particular the ability to vary both carrier concentration and the hopping phase over wide ranges in situ, offers promise of a detailed comparison to theory. Displacement field-tunable metal insulator transitions with interesting precursor phenomena have been reported \cite{Tang:2020aa,ghiotto2021quantum} along with indications of metallic magnetic phases \cite{Tang:2020aa}. The plethora of interesting phases found in our calculations encourage further experimental searches for the stripe and potentially superconducting phases predicted here. In favorable cases stripe phases may be observed via anisotropies in transport measurements although multidomain structures commonly occur and complicate the observations. The spatial modulation of the charge density occurring in a stripe may also be accessible to scanning capacitance probes. Deeper theoretical and experimental understanding of the spin liquid phase that may occur near $U=9t$ at $\phi=0$ and of the anomalous `transition' phases separating the insulating magnet and non-magnetic metal phases are also important open questions.

\begin{acknowledgments}
We would like to thank Edwin M. Stoudenmire and Matthew Fishman for insightful discussions. The DMRG calculations in this manuscript have been performed using the ITensor library~\cite{itensor}. The exact diagonalization calculations have been performed using the Hydra library~\cite{hydra}. J.C., J.Z. and A.J.M acknowledge support from the NSF MRSEC program through the Center for Precision-Assembled Quantum Materials (PAQM) - DMR-2011738. The Flatiron Institute is a division of the Simons Foundation.
\end{acknowledgments}

\appendix

\section{Geometry of the Exact Diagonalization cluster}
\label{sec:edgeometry}
\begin{figure}
    \centering
    \includegraphics[width=\columnwidth]{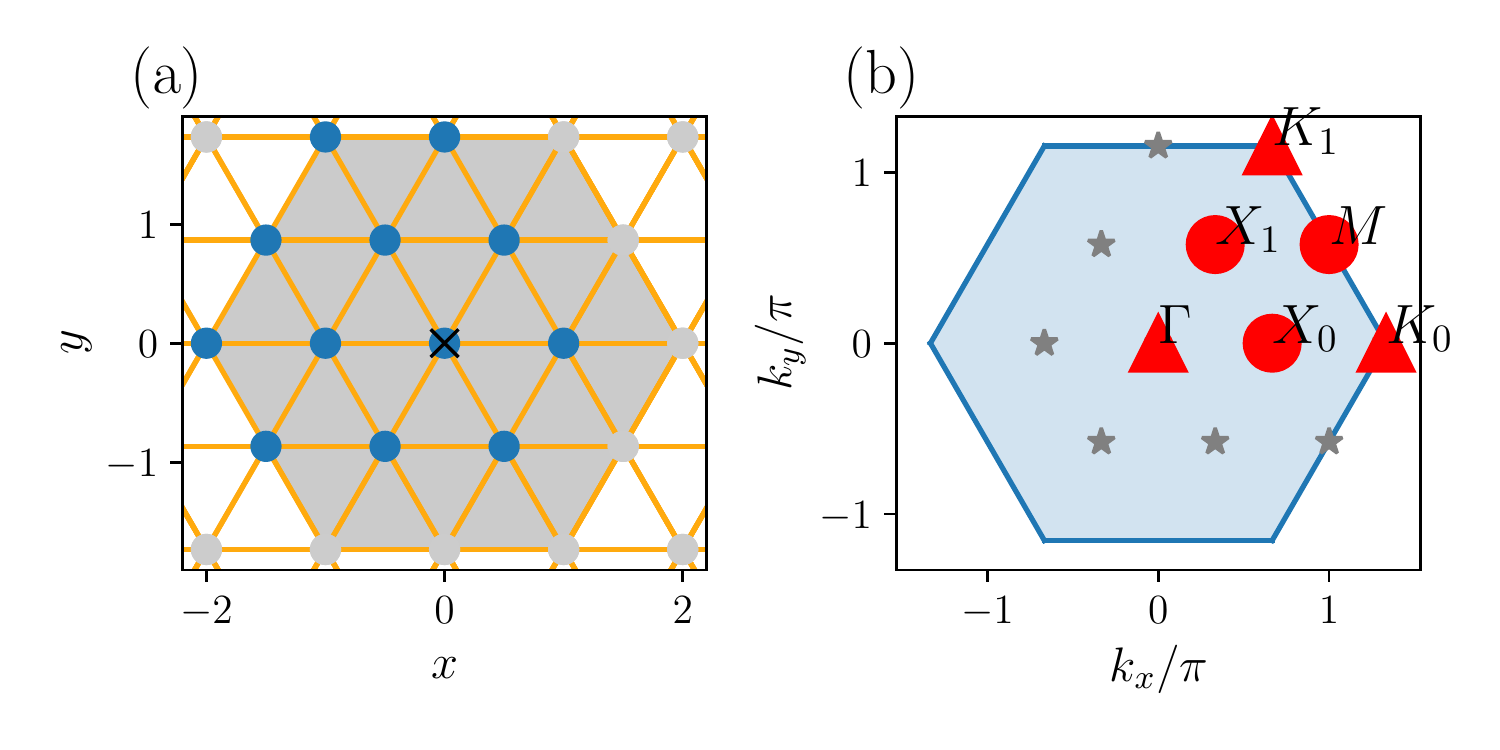}
    \caption{(a) Wigner-Seitz cell of the $N_s=12$ simulation cluster used for exact diagonalization. (b) Momentum resolution in reciprocal space of the $N_s=12$ cluster. This cluster features the highly symmetric $K$ and $M$ points.}
    \label{fig:geo_bz}
\end{figure}

The geometry of the $N_s=12$ site simulation cluster used for the exact diagonalization calculations is shown in \cref{fig:geo_bz}(a). The cluster features the full six-fold rotational symmetry and a mirror reflection symmetry. In reciprocal space it resolves the high symmetry momenta $K$, $M$ and $X$ as shown in \cref{fig:geo_bz}(b). The \mr Hubbard model at $\phi\neq 0$ is only three-fold rotationally symmetric but retains the mirror reflection symmetry. In \cref{fig:first_exc} we show the irreducible representation (irrep) of the first excited state as a function of both $U/t$ and $\phi$. The irreps are labeled by first their momentum quantum number $\bm{k}$ and then their point group representation $\rho$, e.g. $K_0$.A2 refers to the state where $\bm{k} = K_0$ and $\rho = A2$. The point group irreps are denoted by the standardized Mulliken notation~\cite{Mulliken1955}. The momenta $\bm{k} = \Gamma$ and  $\bm{k} = K_0, K_1$ have the little group D3 (dihedral group of order six), whereas the momenta $\bm{k} = M$ and momenta $\bm{k} = X_0, X_1$ have the little group C2. We list the character table of these groups in \cref{tab:irreps}.

\begin{table}[]
\begin{tabular}{|l|c|c|c|}
\hline
 D3  & $I$ & $C_3$ & $C_{3}R$ \\ \hline\hline
A1 & 1 & 1  & 1   \\
A2 & 1 & 1  & -1  \\
E  & 2 & -1 & 0   \\ 
\hline
\end{tabular} \hspace{2cm}
\begin{tabular}{|l|c|c|}
\hline
 C2  & $I$ & $R$ \\ \hline\hline
A & 1 & 1   \\
B & 1 & -1  \\
\hline
\end{tabular}
\caption{Character table for the irreducible representations of the dihedral group D3 (left) and the cyclic group C2 (right).}
\label{tab:irreps}
\end{table}

\bibliography{main.bib}

\end{document}